\documentclass[
 reprint,
 amsmath,amssymb,
 aps,pra,floatfix
]{revtex4-1}

\usepackage{amssymb}
\usepackage{graphicx} 
\usepackage{dcolumn}          
\usepackage{bm}                
\usepackage[normalem]{ulem}
\usepackage{amsmath}
\usepackage{subfigure} 
\usepackage{soul,xcolor}
\usepackage{microtype}

\DeclareMathOperator{\MathieuC}{MathieuC}     
\DeclareMathOperator{\MathieuS}{MathieuS}      

\begin{document}

\title{An optical wormhole from hollow disclinations}

\author{Frankbelson dos S. Azevedo}
\email{frfisico@gmail.com}
\affiliation{Departamento de F\'{i}sica, Universidade Federal Rural de Pernambuco, 
52171-900, Recife, PE, Brazil}

\author{Jos\'{e} Di\^{e}go M. de Lima}
\email{diego.mlima@ufrpe.br}
\affiliation{Departamento de F\'{i}sica, Universidade Federal Rural de Pernambuco, 
52171-900, Recife, PE, Brazil}

\author{Ant\^{o}nio de P\'{a}dua Santos}
\email{antonio.padua@ufrpe.br}
\affiliation{Departamento de F\'{i}sica, Universidade Federal Rural de Pernambuco, 
52171-900, Recife, PE, Brazil}

\author{Fernando Moraes}
\email{fernando.jsmoraes@ufrpe.br}
\affiliation{Departamento de F\'{i}sica, Universidade Federal Rural de Pernambuco, 
52171-900, Recife, PE, Brazil}

\date{\today}

\begin{abstract}
We examine the optical properties of two different configurations of an ordered liquid crystal film on a catenoid forming coreless disclinations. We find the effective optical metric from which we obtain the geodesics and wave modes characterizing thus the propagation of light on this surface. We show that the optical metric describes a two-dimensional section of the spacetime of a conical wormhole. 
\end{abstract}

\maketitle

\section{Introduction}
Wormholes \cite{morris1988wormholes} are shortcuts in spacetime that appear in many science fiction stories, providing convenient justifications for faster than light space displacements as well as for time travel. Even though no one has been detected so far, wormholes are scientifically sound objects which have deserved a vast literature (see \cite{lobo2017wormholes} and references therein). Wormhole-inspired devices are useful not only as experimentally accessible analogues but may have important practical applications as electromagnetic radiation harvesters \cite{ferreira2019optimizing}, for being superabsorbers, or in magnetic resonance imaging \cite{prat2015magnetic}, for letting magnetic fields be transported without distortion. Reference \cite{greenleaf2007electromagnetic} lists a number of possible applications for electromagnetic wormholes built with metamaterials.

A different cosmological object, the cosmic string \cite{vilenkin2000cosmic}, has long been compared to disclinations \cite{de1993physics} in  liquid crystals since they share common formation mechanisms \cite{bowick1994cosmological,mukai2007defect}  and also have similarities in their optical properties  \cite{satiro2005liquid,satiro2008deflection,pereira2011diffraction}. Again, important applications might come out of these analogies, as for instance the optical concentrator in a disclination-based device proposed by two of us and coworkers in Ref. \cite{azevedo2018optical}. Ideal cosmic strings, like ideal disclinations, are problematic due to singularities associated to their respective cores. A curvature singularity in the former and a vector field singularity in the latter. In both cases, more realistic models smooth the singularity over a finite region of space around the center of the defect.

\begin{figure}[htp]
    \centering
    \includegraphics[width=0.8\columnwidth]{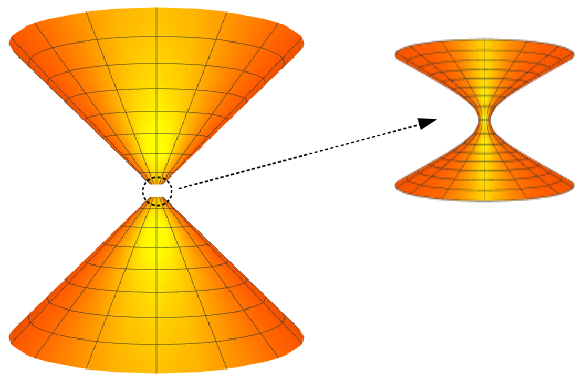}
    \caption{Two-dimensional representation of two cosmic string spacetimes (conical geometry) being joined to form a conic wormhole. }
    \label{cones}
\end{figure}

A way of avoiding the singularity is to cut it out and heal the cut with a graft, imitating the process of ``construction'' of horizon-free wormholes \cite{visser1989traversable}: take two Schwarzschild spacetimes, cut out a four-dimensional region of radius larger than the Schwarzschild radius in each of them, and graft them together at the edges of the cuts. One such construction, the Morris-Thorne wormhole \cite{morris1988wormholes}, became notorious, not only for its simplicity, but also for its relation to Carl Sagan's novel \textit{Contact}. Another such construction \cite{aros1997wormhole}, repeats the process with a cylindrical portion of the conical spacetime surrounding a cosmic string. A two-dimensional representation of the process is shown in Fig. \ref{cones}. The result is two asymptotically conical spacetimes joined by a singularity-free wormhole. That is, in each of these, far from the wormhole mouth, the spacetime is indistinguishable from the one of a cosmic string. Moreover, the string core singularity is removed, its surroundings becoming the throat of the wormhole. It is this idea that we pursue in this article with liquid crystals. We propose a simple way of obtaining a coreless disclination, avoiding the instability problems associated to the director field singularity, and at the same time providing a gravitational wormhole analogue model. We explore our model by studying both geometric and wave optics, finding qualitative agreement with known wormhole results.

In the next sections of this work, nematic liquid crystals are used in a simple device whose geometry generates coreless (or hollow) disclinations representing the junction of two conical spacetimes. In section \ref{section1}, we find the optical metric for two different types of molecular arrangement making disclinations. We then solve the geodesic equation (section \ref{trajectories}) and wave equation (section \ref{waveequation}) for both cases in order to have a clear picture of the propagation of light in these cases. In Section \ref{embedsection} we digress on embedding diagrams and conical metrics. Finally, in section \ref{conclusions} we present our conclusions.

\section{A Liquid Crystal device}
\label{section1}
\begin{figure}[htp]
\centering
    \includegraphics[width=0.49\columnwidth]{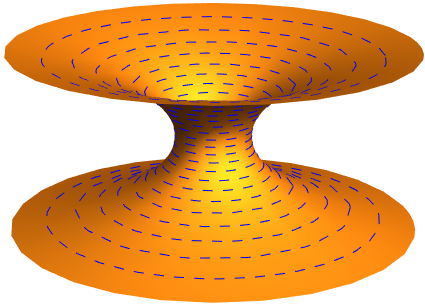}
      \includegraphics[width=0.49\columnwidth]{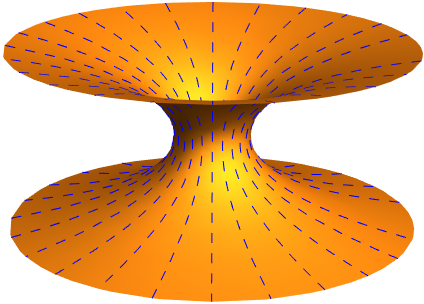}
     \caption{Director field for circular and radial $+ 1$ disclinations on the catenoid, respectively.}
    \label{catenoiddisclination}
\end{figure}

\begin{figure}[htp]
\centering
    \includegraphics[width=0.49\columnwidth]{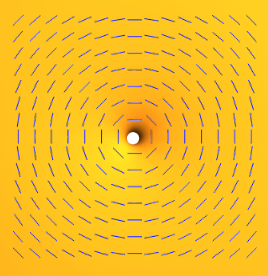}
      \includegraphics[width=0.49\columnwidth]{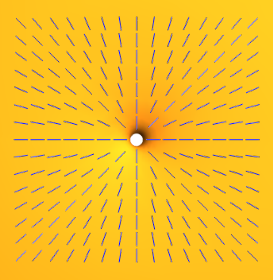}
     \caption{Director field for circular and radial $+ 1$ disclinations on the plane, respectively.}
    \label{flatdisclination}
\end{figure}

 Thin films of nematic liquid crystals on curved surfaces lead to very interesting consequences like defects induced by geometrical frustration \cite{lopez2011frustrated}, for instance. A recent study of the curvature effects on topological defects on such thin films \cite{mesarec2017curvature} demonstrated the possibility of manipulation and control of topological defects by use of curvature. In particular, the authors of Ref. \cite{mesarec2017curvature} focused on nematic ordering on a catenoid and defects with topological charge (winding number) $\pm 1/2$. Borrowing their idea, we study here two versions of a $+1$ winding number disclination, also on a catenoid, which naturally stabilizes and locks the defect around its neck. Moreover, the defects are hollow or coreless due to the catenoid topology. The director field lines corresponding to either defect are, respectively, the circles and catenaries seen in Fig. \ref{catenoiddisclination}. Far away from the catenoid throat, since the surface is asymptotically flat, the defects will look like ordinary $+1$ disclinations on a plane with a hole, like in Fig. \ref{flatdisclination}.

 In what follows we briefly describe how to obtain the optical metric that describes the propagation of light along the nematic film on the catenoid. The ray index associated with the energy propagation in the liquid crystal medium is given by \cite{kleman2007soft}
\begin{equation}
    N^{2} = n_{o}^{2}\cos^{2}\beta + n_{e}^{2}\sin^{2}\beta,
    \label{refrac}
\end{equation}
where $n_{o}$ and $n_{e}$ are the ordinary and extraordinary indices, respectively; $\beta$ is the angle between the tangent vector to the light path (Poynting vector) and the director field line, such that
\begin{equation}
    \cos\beta = \vec{T}\cdot\vec{n},
    \label{beta}
\end{equation}
where $\vec{n}$ is the unitary director vector. The tangent vector to the curve $\vec{R}(\ell)$, that represents the light trajectory, is written as $\vec{T} = \frac{d\vec{R}}{d\ell}$, where $\ell$ is a parameter along the curve. In order to obtain the optical metric, we use the interpretation
\begin{equation}
    ds^{2} = N^{2} d\ell^{2},
    \label{inter}
\end{equation}
where $ds^{2}$ is the line element in Riemannian geometry, which gives the light paths as geodesics in this space \cite{born2013principles}. This is obviously equivalent to Fermat's principle since either geodesic or light path is required to obey the variational principle.

Now, we apply the above relations to a catenoid with its axis along $z$. For this surface, we can use the parametric equations \cite{krivoshapko2015encyclopedia}
\begin{equation}
    \begin{split}
        x & = b_{0}\cosh(z/ b_{0})\cos\phi, \\
        y & = b_{0}\cosh(z/ b_{0})\sin\phi, \\
        z & = z, 
    \end{split}
    \label{cor}
\end{equation}
where $ b_{0}$ is the throat radius and $\phi \in \left[0, 2\pi \right]$. We can reparametrize equations \eqref{cor} in terms of the arc length of the catenary, $\tau = b_{0}\sinh(z/ b_{0})$, 
which results in the catenoid parametrized as 
\begin{equation}
    \begin{split}
        x & = ({\tau^{2}} + b_{0}^{2})^{1/2} \cos\phi, \\
        y & = ({\tau^{2}} + b_{0}^{2})^{1/2} \sin\phi, \\
        z & = b_{0}\sinh^{-1}(\tau/ b_{0}). 
    \label{cors}
    \end{split}
\end{equation}

In Cartesian coordinates, the vector $\vec{R}(\ell)$ can be written as
\begin{equation}
    \vec{R} = x\hat{i} + y\hat{j} + z\hat{k}
\end{equation}
which, according to equations \eqref{cors}, gives 
\begin{equation}
    \vec{R} = (b_{0}^{2} + {\tau^{2}})^{1/2} (\cos\phi \hat{i} + \sin\phi \hat{j}) + b_{0}\sinh^{-1}(\tau/ b_{0}) \hat{k}
\end{equation}
for the geometry of the catenoid. By writing 
\begin{equation}
    d\ell^{2} = dx^{2} + dy^{2} + dz^{2}
\end{equation}
and using equations \eqref{cors}, we get 
\begin{equation}
    \begin{split}
        dx & = \frac{\tau\cos\phi }{(\tau^{2} + b_{0}^{2})^{1/2}}d\tau - (\tau^{2} + b_{0}^{2})^{1/2} \sin\phi d\phi, \\
        dy & = \frac{\tau\sin\phi }{(\tau^{2} + b_{0}^{2})^{1/2}}d\tau + (\tau^{2} + b_{0}^{2})^{1/2} \cos\phi d\phi, \\
        dz & = \frac{ b_{0}}{(\tau^{2} + b_{0}^{2})^{1/2}} d\tau,
    \end{split}
    \label{dx} 
\end{equation}
and then 
\begin{equation}
    d\ell^{2} = d\tau^{2} + (\tau^{2} + b_{0}^{2}) d\phi^{2}.
    \label{dl-metric}
\end{equation}

In addition, we can write
\begin{equation}
    \vec{T} = \frac{dx}{d\ell} \hat{i} + \frac{dy}{d\ell} \hat{j} + \frac{dz}{d\ell} \hat{k},
\end{equation}
which according to equations \eqref{dx} gives us
\begin{equation}
    \begin{split}
        \vec{T} & = \left[\frac{\tau \cos\phi}{(\tau^{2} + b_{0}^{2})^{1/2}} \frac{d\tau}{d\ell} - (\tau^{2} + b_{0}^{2})^{1/2} \sin\phi \frac{d\phi}{d\ell} \right] \hat{i} \\
         & + \left[\frac{\tau \sin\phi}{(\tau^{2} + b_{0}^{2})^{1/2}} \frac{d\tau}{d\ell} + (\tau^{2} + b_{0}^{2})^{1/2} \cos\phi \frac{d\phi}{d\ell}\right] \hat{j} \\
         & + \frac{ b_{0}}{(\tau^{2} + b_{0}^{2})^{1/2}} \frac{d\tau} {d\ell} \hat{k}. 
    \label{tang}
    \end{split}
\end{equation}
Now, by specifying  the director $\vec{n}$ configuration, we can study the two different types of molecular arrangement depicted in Fig. \ref{catenoiddisclination}, circular and radial (along the catenary lines).

\subsection{Optical metric for circular disclination on the catenoid}
\label{firstLC}
For the nematic liquid crystal molecules circularly distributed on the catenoid (see Fig. \ref{catenoiddisclination}), the director is given by
\begin{equation}
    \vec{n} = -\sin\phi \hat{i} + \cos \phi \hat{j}.
    \label{arr1}
\end{equation}
The angle between the tangent vector and the director is obtained from equations \eqref{beta}, \eqref{tang}, and \eqref{arr1} as
\begin{equation}
    \cos^{2}\beta = (\tau^{2} + b_{0}^{2})\left(\frac{d\phi}{d\ell}\right)^{2}.
\end{equation}
Therefore, by using equation \eqref{refrac}, we find 
\begin{equation}
    N^{2} = (n_{o}^{2} - n_{e}^{2})(\tau^{2} + b_{0}^{2})\left(\frac{d\phi}{d\ell}\right)^{2} + n_{e}^{2}.
\end{equation}
Hence, by using equation \eqref{inter}, we get
\begin{equation}
    ds^{2} = n_{e}^{2} [d\tau^{2} + \alpha^{2}(\tau^{2} + b_{0}^{2}) d\phi^{2}],
\end{equation}
where 
\begin{equation}
   \alpha = {n_{o}}/{n_{e}} . \label{alpha}
\end{equation}
 Rescaling $ds \rightarrow {ds}/{n_{e}}$, we obtain a new metric,
\begin{equation}
    ds^{2} = d\tau^{2} + \alpha^{2} (\tau^{2} + b_{0}^{2}) d\phi^{2},
    \label{metric1.1}
\end{equation}
which is analogous to equation \eqref{dl-metric} with the extra factor $\alpha$ which brings conical features to this effective geometry as will be seen in Section \ref{embedsection}.

\subsection{Optical metric for radial disclination on the catenoid}
\label{secondLC}
For the nematic liquid crystal configuration where the molecules is along the catenary lines (see Fig. \ref{catenoiddisclination}) we have
\begin{equation}
\begin{split}
    \vec{n} = \frac{d} {d\tau} \bigg[\left({\tau^{2}} + b_{0}^{2} \right)^{1/2}(\cos\phi_{0} & \hat{i} + \sin\phi_{0} \hat{j}) 
    \\ & + b_{0}\sinh^{-1} (\tau/ b_{0}) \hat{k} \bigg], 
\end{split}
\end{equation}
for a given $\phi=\phi_{0}$. Thus, we get
\begin{equation}
    \vec{n} = \dfrac{1}{{\left({\tau^{2}} + b_{0}^{2} \right)^{1/2}}} \left[\tau(\cos\phi_{0} \hat{i} + \sin\phi_{0} \hat{j}) + { b_{0}}\hat{k}\right].
    \label{arr2}
\end{equation}
Hence, from equations \eqref{beta}, \eqref{tang}, and \eqref{arr2} we obtain 
\begin{equation}
    \cos^{2}\beta = \left(\frac{d\tau}{d\ell}\right)^{2},
\end{equation}
and then the refractive index is given by
\begin{equation}
    N^{2} = \frac{{n_{o}^{2} - n_{e}^{2}}}{({\tau^{2}} + b_{0}^{2})^{2}} {({\tau^{2}}\cos\phi + b_{0}^{2})^{2}} + n_{e}^{2}.
\end{equation}
Therefore, from equation \eqref{inter} we find
\begin{equation}
    ds^{2} = n_{o}^{2} [d\tau^{2} + \alpha^{-2}(\tau^{2} + b_{0}^{2}) d\phi^{2}],
\end{equation}
where $\alpha$ is given by Eq. \eqref{alpha}. Rescaling $ds \rightarrow {ds}/{n_{o}}$, we obtain  
\begin{equation}
    ds^{2} = d\tau^{2} + \alpha^{-2}(\tau^{2} + b_{0}^{2}) d\phi^{2}, \label{metric2}
\end{equation}
analogous to \eqref{metric1.1} but now with inverted $\alpha$. Both metrics represent then  effective conical wormhole geometries.

Since for nematic liquid crystals composed of elongated molecules, typically $n_{o} < n_{e}$ (optically positive nematic) \cite{kleman2007soft}, then $\alpha < 1$. {For $\tau \gg b_{0}$ (far from the throat), the metrics \eqref{metric1.1} and \eqref{metric2} approximate the optical metrics of a $+ 1$ disclination in nematics \cite{satiro2006lensing,satiro2008deflection}} in the circular and radial configurations, respectively. The isotropic case (no defect) is obtained when $n_{o} = n_{e}$, or $\alpha = 1$. In this case, the metrics \eqref{metric1.1} and \eqref{metric2} reduce to the one of the Morris-Thorne wormhole $t = \text{const.}$, $\theta = \pi/2$ section, embedded in 3D Euclidean space \cite{morris1988wormholes}. For $0 < \alpha < 1$ the metrics \eqref{metric1.1} and \eqref{metric2} are similar to the ones of an asymptotically conical  wormhole with a global monopole charge \cite{jusufi2018conical} and of cosmic string wormholes \cite{aros1997wormhole, eiroa2004cylindrical}. This will be discussed in detail in Section \ref{embedsection}.

\section{Ray optics in the effective conical wormhole geometry}
\label{trajectories}
From this point on, we examine the light propagation in the geometric background given by the metrics \eqref{metric1.1} and \eqref{metric2} with $0<\alpha < 1$. We start studying the geodesic equation in order to find the light ray trajectories. Since both metrics are formally the same except for a multiplicative factor in their angular part, we perform the calculations with \eqref{metric1.1}. The results for metric \eqref{metric2} are immediately obtained by replacing $\alpha \to \alpha^{-1}$.

To find the geodesics we follow Ref. \cite{muller2008exact}, where the author determines null and timelike geodesics in terms of elliptic integrals. We shall develop the same procedure considering the angular deficit/surplus factor $\alpha$. We take the Lagrangian as
\begin{equation}
  \mathcal{L} = g_{\mu\nu} \frac{d x^{\mu}}{d\lambda} \frac{d x^{\nu}}{d\lambda},
\end{equation}
where the geodesics are obtained as solutions of Euler-Lagrange equation:
\begin{equation}
    \frac{\partial \mathcal{L}}{\partial x^{\mu}} - \frac{d}{d \lambda}\bigg(\frac{\partial\mathcal{L}}{\partial \dot{x}^{\mu}}\bigg) = 0.
\end{equation}
The Lagrangian must obey  $\mathcal{L} = \kappa c^{2}$, where $\kappa = 0$ for lightlike geodesics, $\kappa = - 1$ for timelike geodesics, and $c$ is the speed of light in vacuum which, from here on, we consider to be unitary. From the line element in equation \eqref{metric1.1}, we can write the Lagrangian as 
\begin{equation}
    \mathcal{L} = - \dot{t}^{2} + \dot{\tau}^{2} + \alpha^{2}(\tau^{2} + b_{0}^{2})\dot{\phi}^{2}.
    \label{lagrangian}
\end{equation}

\begin{figure}[htp!]
	\centering
 	\includegraphics[width=0.8\columnwidth]{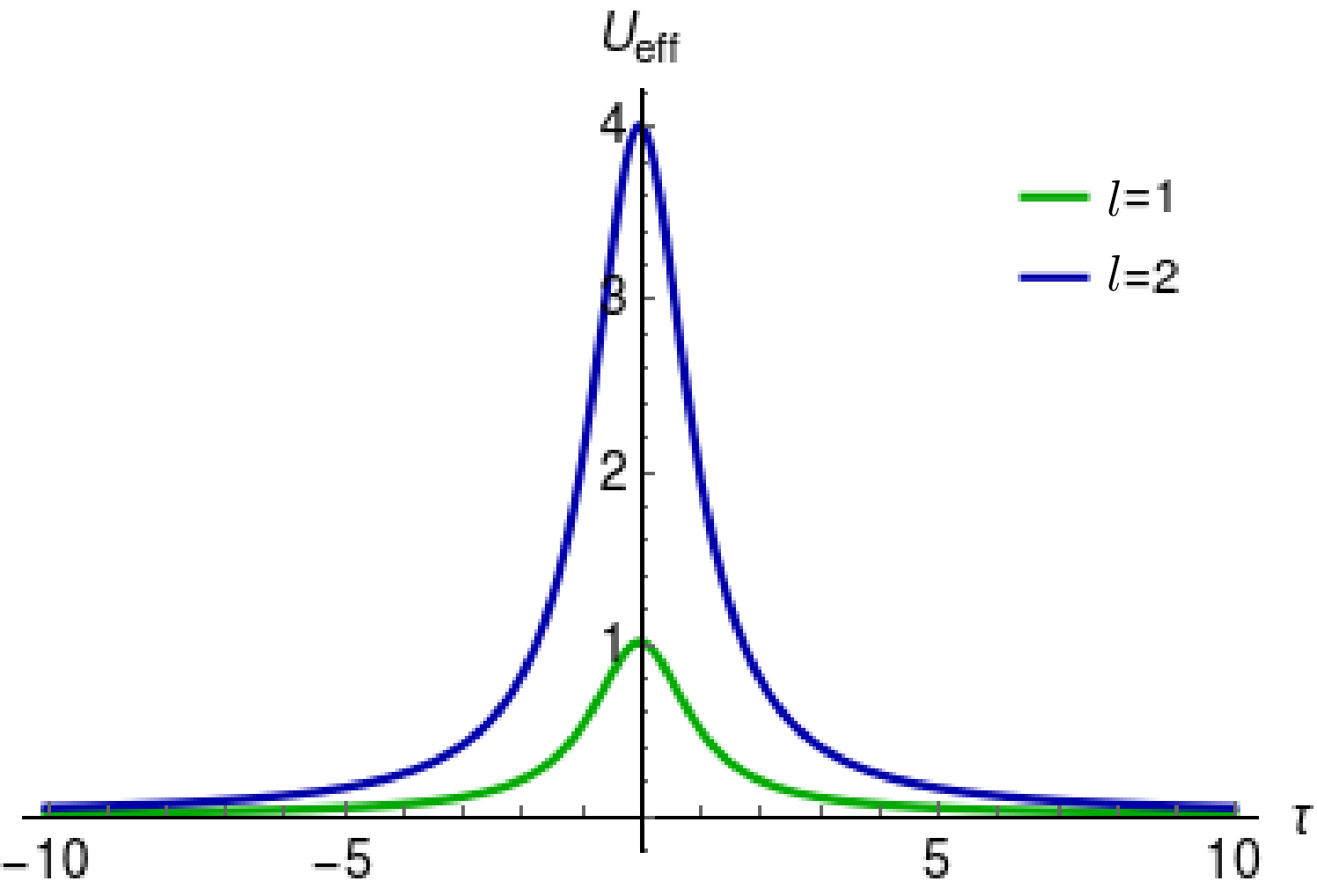}
	\caption{Effective potential for $l = 1$ and $l = 2$. We consider here $b_{0} = 1$ and $\kappa = 0$ for light rays.}
	\label{class_Ueff}
\end{figure}

 Now, let us define a new angular variable $\varphi = \alpha\phi$. The Lagrangian \eqref{lagrangian} becomes
\begin{equation}
    \mathcal{L} = - \dot{t}^{2} + \dot{\tau}^{2} + (\tau^{2} + b_{0}^{2})\dot{\varphi}^{2}.
    \label{lagrangian2}
\end{equation}
Since $\mathcal{L}$ does not depend explicitly on $t$ and $\varphi$, $\frac{\partial \mathcal{L}}{\partial \dot{t}}$ and $\frac{\partial \mathcal{L}}{\partial \dot{\varphi}}$ are constants of motion, that we label $2K$ and $2l$, respectively. It follows that
\begin{equation}
    K = \dot{t} \quad \text{and} \quad l = (\tau^{2} + b_{0}^{2})\dot{\varphi},
    \label{fi_ponto}
\end{equation}
such that we get the differential equation 
\begin{equation}
    \dot{\tau}^{2} = \kappa + K^{2} - \dfrac{l^{2}}{\tau^{2} + b_{0}^{2}}.
    \label{tau_dot}
\end{equation}
From this equation, we can identify the effective potential
\begin{equation}
    U_{eff} = -\kappa + \dfrac{l^{2}}{\tau^{2} + b_{0}^{2}}.\label{class_Ueff2}
\end{equation}
 The behavior of the effective potential $U_{eff}$ is plotted in Fig. \ref{class_Ueff} for $l = 1$ and $l = 2$, respectively. It is clear that there is no stable equilibrium point.

The orbital motion can be obtained in terms of the angular variable as $\tau = \tau(\varphi)$. In this case, from equation \eqref{tau_dot} and using equation \eqref{fi_ponto}, we get
\begin{equation}
    \left(\dfrac{d\tau}{d\varphi}\right)^{2} = \dfrac{\dot{\tau^{2}}}{\dot{\varphi^{2}}} = \dfrac{(\kappa + K^{2})}{l^{2}}(\tau^{2} + b_{0}^{2})^{2} -(\tau^{2} + b_{0}^{2}). \label{eqorbit}
\end{equation}
Defining 
\begin{equation}
    \rho = \frac{b_{0}}{a}\sqrt{\tau^{2} + b_{0}^{2}} ,
    \label{r}
\end{equation}
and with 
\begin{equation}
    a = \frac{b_{0}}{l}\sqrt{\kappa + K^{2}} ,
    \label{param_a}
\end{equation}
  Eq. \eqref{eqorbit} is changed into
\begin{equation}
    \bigg(\frac{d \rho}{d\varphi}\bigg)^{2} = (1 - a^{2}\rho^{2})(1 -\rho^{2}).
\end{equation}

Considering the original angular variable $\phi$, we have
\begin{equation}
    \phi(\rho) =\phi(\rho_i) \pm \frac{1}{\alpha}\int_{\rho_{i}}^{\rho}\frac{d\rho}{\sqrt{(1 - a^{2}\rho^{2})(1 -\rho^{2})}},
\end{equation}
whose solution is given by
\begin{equation}
    \phi(\rho) =\phi(\rho_i)  \pm \frac{1}{\alpha} [F(\text{sin}^{-1}\rho, a^2) - F(\text{sin}^{-1}\rho_i, a^2)],
    \label{fi}
\end{equation}
where $F$ is the elliptic integral of the first kind \cite{abramowitz1948handbook,gradshteyn1988tables}. Recall that equation \eqref{fi} gives both the trajectories of particles ($\kappa = -1$), as well as light trajectories ($\kappa = 0$) in the conical 2D wormhole geometry of metric \eqref{metric1.1}. By making $\alpha\to\alpha^{-1}$, it gives also the trajectories in the geometry of metric \eqref{metric2}.

\begin{figure*}[htp]
	\centering
	(a)\includegraphics[width=0.475\columnwidth]{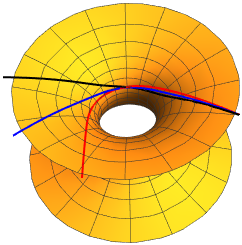}
	\hfill (b) \includegraphics[width=0.525\columnwidth]{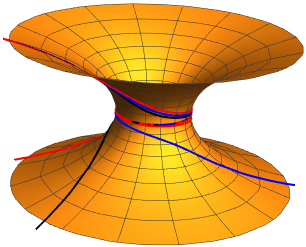}
	 \hfill (c)\includegraphics[width=0.525\columnwidth]{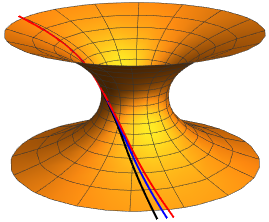}
	\caption{Geodesics for (a) $a < 1$, (b) $a = 1$, and (c) $a > 1$. The blue lines represents the isotropic case $\alpha = 1$. The red and black lines represent, respectively, circular (deficit angle) and radial (surplus angle) disclinations with $\alpha = 0.85$ for (a) and (b), and with $\alpha = 0.98$ for (c).}
	\label{orbs}
\end{figure*}

In Figs. \ref{orbs} (a), \ref{orbs} (b), and \ref{orbs} (c) we plot geodesics for different values of $a$ and $\alpha$. In each case, the geodesics start at the same point and with the same shooting angle. Notice that when the defect parameter $\alpha = 1$, which corresponds to no angular deficit/surplus, the trajectory is due only to the catenoid geometry. The role of $\alpha$ is clear in Fig. \ref{orbs} (a), bending the trajectory away from the $\alpha = 1$ case as if there was less (deficit angle defect) or more (surplus angle defect) available space. We note that, by adjusting $\alpha$, it is possible to minimize or maximize the effects of the catenoid curvature on the geodesics. It is instructive to compare Fig. \ref{orbs} (a) with Fig. 5 of Ref. \cite{satiro2006lensing} which gives a picture of the deflection of light by disclinations. This corresponds to a top view of Fig. \ref{orbs} (a) far from the catenoid mouth. In Fig. \ref{orbs} (b) is a situation where the geodesics wind around the catenoid neck and then go away. Finally, in Fig. \ref{orbs} (c), we present the geodesics for $a > 1$. As Eq. \eqref{class_Ueff2} indicates, $l$ is an effective angular momentum and, from Eq. \eqref{class_Ueff2} we see that $a$ becomes larger as $l$ decreases. The limit case $l = 0$ corresponds to trajectories along catenaries on the surface. Figure \ref{orbs} (c) shows geodesics with low angular momentum thus approaching the shape of catenaries.

\section{Wave optics in the effective conical wormhole geometry}
\label{waveequation}
In the previous section, we obtained the trajectories of light in the effective geometry of a catenoid with a film of oriented nematics. Now, we examine the corresponding propagating wave modes. We start with Helmholtz equation in the effective geometry,
\begin{equation}
    \left(\Delta_{g} + k^{2}\right)\Psi = 0,
    \label{waveeq}
\end{equation}
where $\Delta_{g}$ is the Laplace-Beltrami operator, replacing the usual Laplacian, and $k$ is the wave number \cite{schultheiss2010optics}. For a generic line element $ds^{2} = g_{ij} dx^{i} dx^{j}$, the Laplace-Beltrami operator acting on the scalar function $\Psi$ is given by 
\begin{equation}
    \Delta_g\Psi=\frac{1}{\sqrt{|\text{det}(g_{ij})|}}\partial_i \left( \sqrt{|\text{det}(g_{ij})|}g^{ij}\partial_j \Psi\right).
\end{equation}
Keeping in mind that we can go from the circular to the radial disclination by making $\alpha\to\alpha^{-1}$, and using metric \eqref{metric1.1} we get \cite{piropo2019surfing} 
\begin{equation}
    \Delta_{g} = \dfrac{\partial^{2}}{\partial \tau^{2}} + \dfrac{\tau}{\tau^{2} + b_{0}^{2}}\dfrac{\partial}{\partial\tau} + \dfrac{1}{\alpha^{2} (\tau^{2} + b_{0}^{2})}\dfrac{\partial^{2}}{\partial\phi^{2}},
\end{equation}
for the catenoid with the circular disclination.
From equation \eqref{waveeq} we obtain then
\begin{equation}
    \dfrac{\partial^{2}\Psi}{\partial \tau^{2}} + \dfrac{\tau}{\tau^{2} + b_{0}^{2}}\dfrac{\partial\Psi}{\partial\tau} + \dfrac{1}{\alpha^{2} (\tau^{2} + b_{0}^{2})}\dfrac{\partial^{2}\Psi}{\partial\phi^{2}} + k^{2}\Psi = 0.
\end{equation}
Using the ansatz $\Psi(\tau,\phi) = e^{im\phi}Z(\tau)$, where $m = 0,\pm 1,\pm 2,\dots$ due to the periodic boundary condition on $\phi$, it follows that
\begin{equation}
    \dfrac{d^{2}Z}{d\tau^{2}} + \dfrac{\tau}{\tau^{2} + b_{0}^{2}}\dfrac{dZ}{d\tau} + \left[k^{2} - \dfrac{m^{2}}{\alpha^{2}(\tau^{2} + b_{0}^{2})}\right]Z = 0.
    \label{tau-catenoid}
\end{equation}

\begin{figure}[htp!]
    \centering 
     (a) \includegraphics[width=0.8\columnwidth]{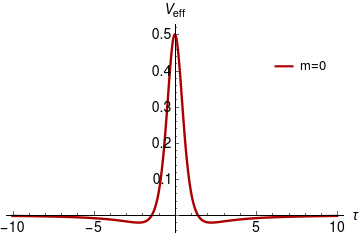} \\
     (b) \includegraphics[width=0.8\columnwidth]{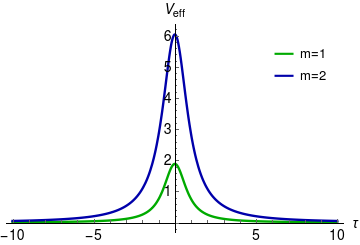} 
    \caption{Effective potential for the catenoid with a circular disclination, for a few values of $m$ and $\alpha = 0.85$. In graph (a), for $m = 0$, there are two minima at $\tau = \pm\sqrt{5}$ (we are setting $b_{0} = 1$) and a maximum at $\tau = 0$. In graph (b), for $m = 1$ and $m = 2$, the potential has only a maximum at $\tau = 0$. For the three curves, the potential goes to zero asymptotically and has successive increasing values of the barrier with increasing $m$. The shape of the potential is similar for the radial disclination case (not plotted here).}
    \label{fig:effpot}
\end{figure}

Following the work of Kar et al. \cite{kar1994scalar} on scalar waves in a wormhole geometry in $2 + 1$ dimensions, we can write equation \eqref{tau-catenoid} in reduced form with the change $Z(\tau)=(\tau^{2} + b_{0}^{2})^{-1/4}\chi(\tau),$ resulting in an effective one-dimensional Schr\"odinger-like equation
\begin{equation}
    \dfrac{d^{2}\chi}{d\tau^{2}}+\left[k^{2} - V_{eff}(\tau) \right]\chi = 0,
    \label{reducedform}
\end{equation}
with the effective potential (see Fig. \ref{fig:effpot})
\begin{equation}
    V_{eff}(\tau) = \dfrac{2 b_{0}^{2} - \tau^{2}}{4(\tau^{2} + b_{0}^{2})^{2}} + \dfrac{m^{2}}{\alpha^{2}(\tau^{2} + b_{0}^{2})}.
    \label{effective}
\end{equation}
The second term in Eq. \eqref{effective} is a centrifugal term for the effective angular momentum $m^{2}/\alpha^{2}$. As we can see, this potential goes to zero when $\tau \rightarrow \pm \infty,$ has a maximum at $\tau = 0$ for any values of $m$ and $\alpha$, and has minima at
\begin{equation}
    \tau = \pm \left(\dfrac{5\alpha^{2} + 4m^{2}}{\alpha^{2} - 4m^{2}}\right)^{1/2} b_{0},
\end{equation}
for $|m| < \alpha/2$ (circular disclination) or $|m| < 1/2\alpha $ (radial disclination). The effect of the liquid crystal is evident here: the $\alpha/2$ (or $1/2\alpha$) factor determines for which values of $m$ the effective potential has attractive regions. For the values that we use here, the effective potential $V_{eff}(\tau)$ has attractive regions only for $m = 0$, as shown in Fig. \ref{fig:effpot}.

Turning our attention again to equation \eqref{tau-catenoid}, we obtain for $\tau \gg b_{0}$ 
\begin{equation}
    \tau^{2}\dfrac{d^{2}Z}{d \tau^{2}} + \tau \dfrac{dZ}{d \tau} + \left(k^{2} \tau^{2} - \dfrac{m^{2}}{\alpha^{2}}\right)Z = 0,
    \label{bessel}
\end{equation}
that is, when far from the throat the wave equation is a Bessel equation. On the other hand, we get for $\tau \ll b_{0}$
\begin{equation}
  \dfrac{d^{2}Z}{d \tau^{2}} + \left( k^{2}-\dfrac{m^{2}}{\alpha^{2} b_{0}^{2}}\right)Z = 0.
  \label{trivialeq}
\end{equation}
Thus, equation \eqref{bessel} is equivalent to the radial part of Helmholtz equation in polar coordinates, which agrees with the fact that far from the throat the wormhole spacetime is essentially flat. This behavior is in agreement with Ref. \cite{pereira2011diffraction}. Near the throat, from \eqref{trivialeq}, light perceives (approximately) the metric of a cylinder and a repulsive centrifugal potential, $\frac{m^{2}}{\alpha^{2} b_{0}^{2}}$. This is indeed expected from either metric \eqref{metric1.1} or \eqref{metric2}: as $\tau \gg b_{0}$ we have  metric that approximates the one of a cosmic string (liquid crystal disclination), and as $\tau \ll b_{0}$ we have an approximate metric with constant coefficients describing a cylinder.

Now, if we take $\tau = b_{0}\sinh\left({z}/{b_{0}}\right)$ (arc length of the catenary) in equation \eqref{tau-catenoid}, we get in terms of $z$
\begin{equation}
    \dfrac{d^{2}Z}{dz^{2}} + \left[k^{2}\cosh^{2}\left({z}/{ b_{0}}\right) - \dfrac{m^{2}}{\alpha^{2}{ b_{0}^{2}}}\right]Z = 0.
    \label{ztilde-eq}
\end{equation}
Introducing the dimensionless quantity $\Tilde{z}={z}/{ b_{0}}$, we get from \eqref{ztilde-eq}, after some algebraic manipulations,
\begin{equation}
    \dfrac{d^{2}Z}{d{\Tilde{z}}^{2}}+\left[2q\cosh\left({2\Tilde{z}}\right)-\epsilon\right]Z = 0,
    \label{Mathieu}
\end{equation}
which is known as the modified Mathieu equation \cite{abramowitz1948handbook} with parameters $\epsilon = \frac{m^{2}}{\alpha^{2}} - \frac{k^{2} b_{0}^{2}}{2}$ and $q = \frac{k^{2} b_{0}^{2}}{4}$. We point out that equation \eqref{Mathieu} has Bessel functions as asymptotic solutions. For $\alpha = 1$, an equivalent equation was found in \cite{kar1994scalar}. In terms of $\tilde{z}$, the even solution for \eqref{Mathieu} is
\begin{equation}
    Z_{e}(\tilde{z}) = A_{1}\MathieuC\left(\epsilon,q,{i\Tilde{z}} \right)
    \label{mathieueven}
\end{equation}
and the odd solution is
\begin{equation}
    Z_{o}(\tilde{z}) = -iA_{2}\MathieuS\left(\epsilon,q,{i\Tilde{z}}\right),
    \label{mathieuodd}
\end{equation}
with $\MathieuC$ and $\MathieuS$ being the Mathieu cosine and sine functions \cite{gutierrez2003mathieu,zaitsev2002handbook,McLachlan}, respectively, and where $A_{1}$ and $A_{2}$ are multiplicative constants (amplitudes).

For $q \ll 1$ (small values of $k^{2} b_{0}^{2}$), the solutions in \eqref{mathieueven} and \eqref{mathieuodd} can be expressed in terms of hyperbolic functions. At the limit when $q \rightarrow 0$, we get $Z_{e}\sim A_{1} \cosh\left(\frac{m\Tilde{z}}{\alpha}\right )$ and $Z_{o} \sim A_{2}\frac{\alpha}{m}\sinh\left( \frac{m\Tilde{z}}{\alpha} \right)$, so that both lose their oscillatory character. The solutions become considerably more complicated for $q \gg 1$, substantially increasing the computational cost for numerical calculations \cite{abramowitz1948handbook,McLachlan,olver2010nist}. The modified Mathieu functions behavior is complex, particularly due to the dependence of the functions on the parameter $q$ \cite{gutierrez2003mathieu}. The case when $q < 0$ (that is $k^{2} < 0$) is only possible for exotic particles like tachyons \cite{feinberg1967possibility}, which we will not discuss here. We can now draw some graphs to better represent the structure of the wave modes. 

\begin{figure}[htp!]
    \centering 
    (a)\includegraphics[width=0.8\columnwidth]{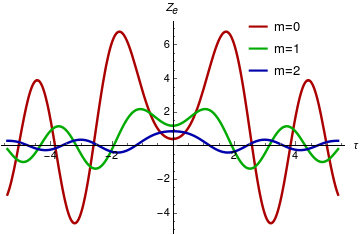} \\
    (b)\includegraphics[width=0.8\columnwidth]{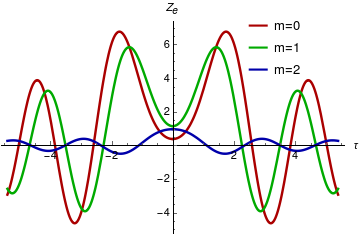}
    \caption{The radial wave modes $Z_{e}$ (even solutions) for $\alpha = 0.85$, $q = 2$, and a few values of $m$. Graph (a) is for the circular disclination and graph (b) is for for the radial disclination. We are using arbitrary multiplicative constants.}
    \label{wav}
\end{figure}

\begin{figure}[htp!]
    \centering 
    (a)\includegraphics[width=0.8\columnwidth]{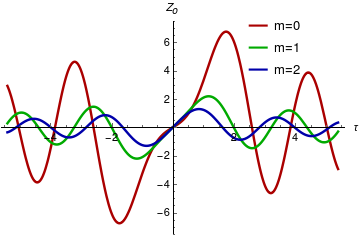} \\
    (b)\includegraphics[width=0.8\columnwidth]{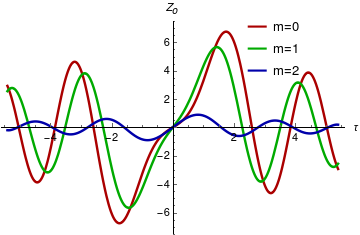}
    \caption{The radial wave modes $Z_{o}$ (odd solutions) for $\alpha = 0.85$, $q = 2$, and a few values of $m$. Graph (a) is for the circular disclination and graph (b) is for for the radial disclination. We are using arbitrary multiplicative constants.}
    \label{wav2}
\end{figure}

\begin{figure*}[htp!]
    \centering 
    \begin{minipage}[c]{1\columnwidth}
    (a)
    \includegraphics[scale=0.2]{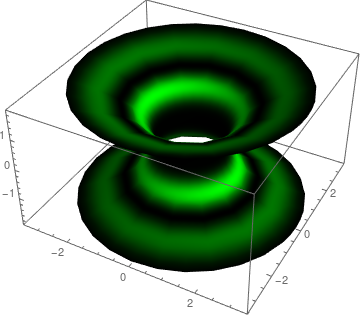}
    \includegraphics[scale=0.21]{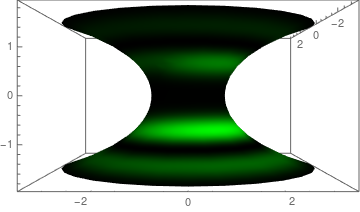}
    \includegraphics[scale=0.17]{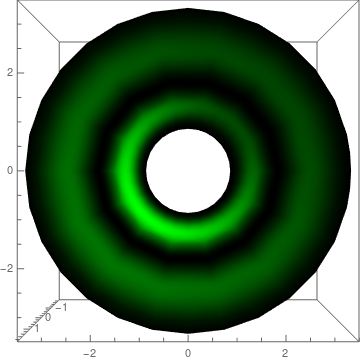}
    \includegraphics[scale=0.2]{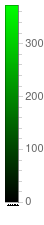} \\
    (b)
    \includegraphics[scale=0.2]{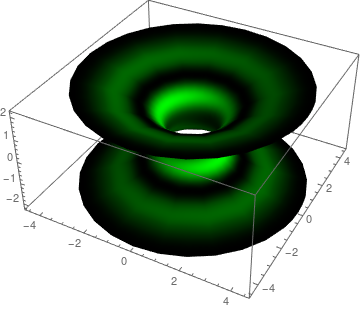}
    \includegraphics[scale=0.21]{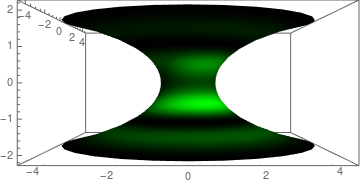}
    \includegraphics[scale=0.17]{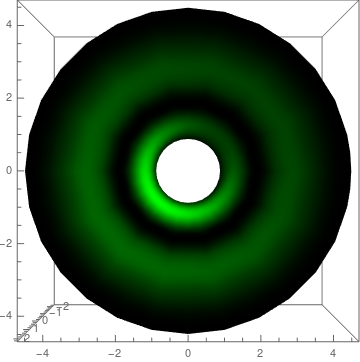}
    \includegraphics[scale=0.2]{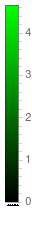} \\
    (c)
    \includegraphics[scale=0.2]{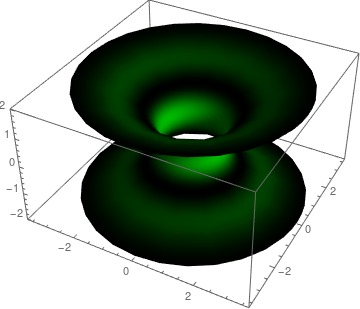}
    \includegraphics[scale=0.2]{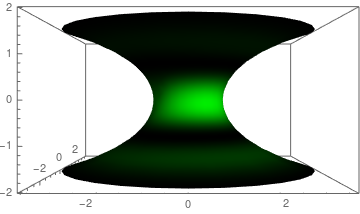}
    \includegraphics[scale=0.195]{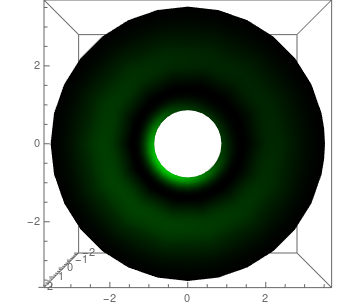}
    \includegraphics[scale=0.2]{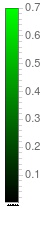} \\
    (d)
    \includegraphics[scale=0.2]{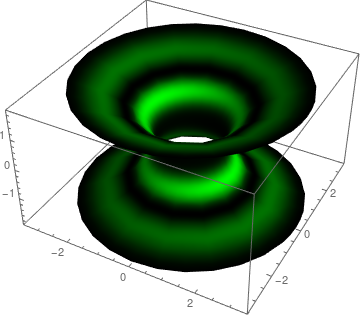}
    \includegraphics[scale=0.21]{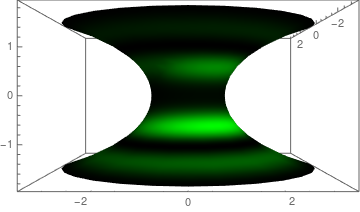}
    \includegraphics[scale=0.17]{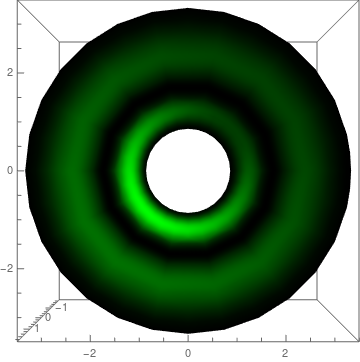}
    \includegraphics[scale=0.2]{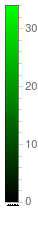} \\
    (e)
    \includegraphics[scale=0.2]{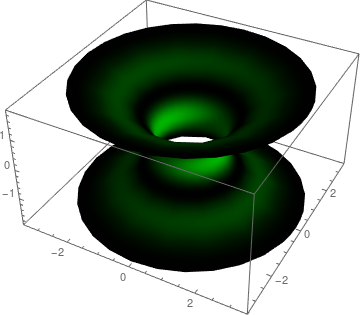}
    \includegraphics[scale=0.21]{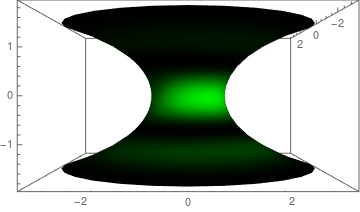}
    \includegraphics[scale=0.17]{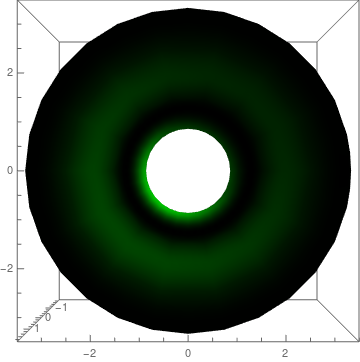}
    \includegraphics[scale=0.2]{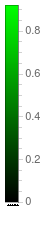}
    \caption{Intensity profiles of the radial wave modes $Z_{e}$ (even solution) on the catenoid surface for $\alpha = 0.85$, $q = 2$ and some values of $m$. (a) For $m = 0$ (in this case both circular and radial disclination cases have the same plot). (b) Circular disclination case for $m = 1$. (c) Circular disclination case for $m = 2$. (d) Radial disclination case for $m = 1$. (e) Radial disclination case for $m = 2$.}
    \label{waves4}
    \end{minipage}
    \hfill
\begin{minipage}[c]{1\columnwidth}
    \centering 
    (a)
    \includegraphics[scale=0.2]{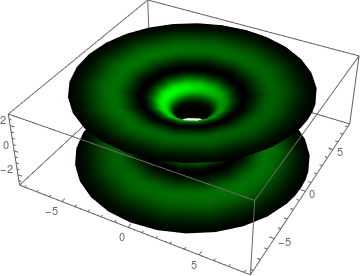}
    \includegraphics[scale=0.21]{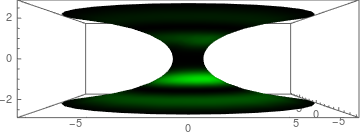}
    \includegraphics[scale=0.16]{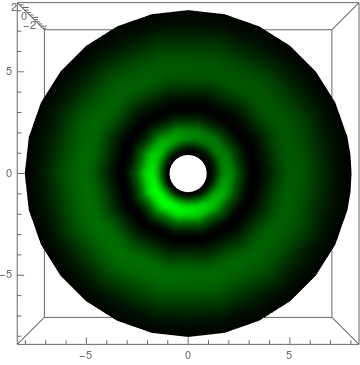}
    \includegraphics[scale=0.2]{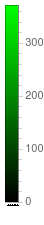}
    (b)
    \includegraphics[scale=0.2]{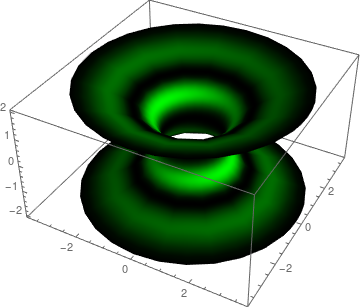}
    \includegraphics[scale=0.2]{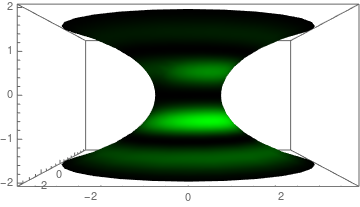}
    \includegraphics[scale=0.19]{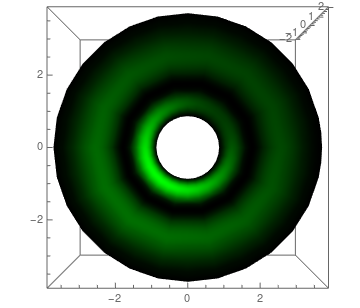}
    \includegraphics[scale=0.2]{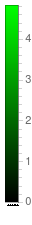}
    (c)
    \includegraphics[scale=0.2]{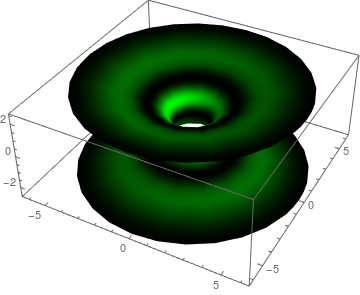}
    \includegraphics[scale=0.195]{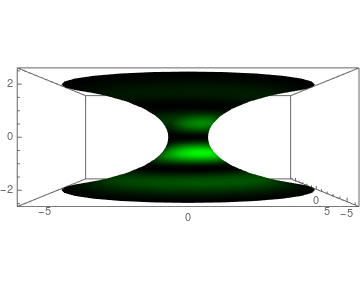}
    \includegraphics[scale=0.195]{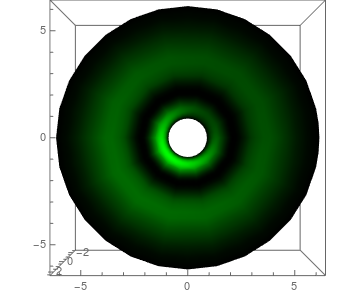}
    \includegraphics[scale=0.2]{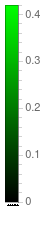}
     (d)
    \includegraphics[scale=0.2]{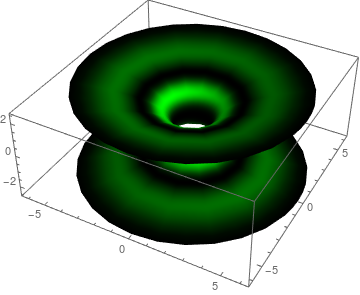}
    \includegraphics[scale=0.2]{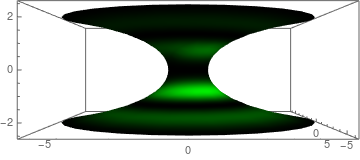}
    \includegraphics[scale=0.165]{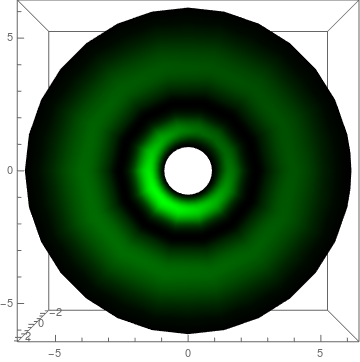}
    \includegraphics[scale=0.2]{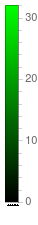}
    (e)
    \includegraphics[scale=0.2]{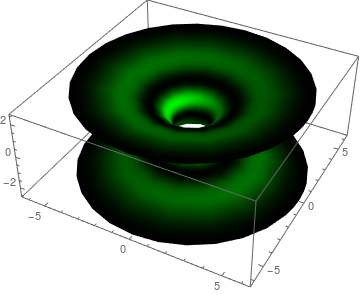}
    \includegraphics[scale=0.195]{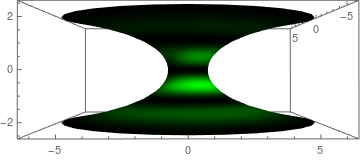}
    \includegraphics[scale=0.19]{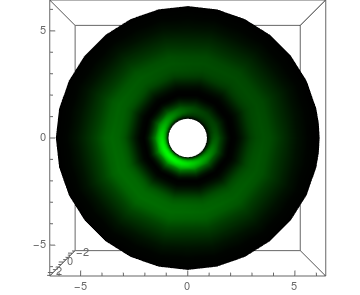}
    \includegraphics[scale=0.2]{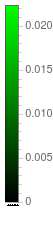}
    \caption{Intensity profiles of the radial wave modes $Z_{o}$ (odd solution) on the catenoid surface for $\alpha = 0.85$, $q = 2$ and some values of $m$. (a) For $m = 0$ (in this case both circular and radial disclination cases have the same plot). (b) Circular disclination case for $m = 1$. (c) Circular disclination case for $m = 2$. (d) Radial disclination case for $m = 1$. (e) Radial disclination case for $m = 2$.}
    \label{waves5}
    \end{minipage}
\end{figure*}

In Fig. \ref{wav} we present the wave modes $Z_{e}$ for a few values of $m$. The incoming waves approach the throat visibly increasing their amplitudes. We can also see that the bigger the value of $m$, the stronger the oscillations become, and the smaller the amplitude is. When we set $m = 1$ or $m = 2$, the amplitude of the wave rapidly decreases as it is close to the throat. In this case, there is a greater probability to localize the wave near the throat, but not exactly at it. For $m = 0$, there is a greater probability to localize the wave at the throat. In Fig. \ref{wav2} is shown the wave modes $Z_{o}$ for different values of $m$. In this case, the incoming waves arrive at the throat increasing their amplitudes and then decreasing to reach null amplitude at $\tau = 0$, since these are the odd solutions. We also see that the bigger the value of $m$, the stronger the oscillations become, and the smaller the amplitude is. There is a greater probability to localize the wave near the throat, but probability zero exactly at the throat. The intensity profiles of the radial wave modes $|Z_{e}|^{2}$ and $|Z_{o}|^{2}$ on the catenoid surface for some values of $m$ and $\alpha=0.85$ are shown in Figs. \ref{waves4} and \ref{waves5}, respectively. 

\section{On embedding diagrams and conical metrics \label{embedsection}}
Wormholes in spacetime are usually represented by ``embedding diagrams'', which are 2D slices of the 4D structure immersed in Euclidean 3D space. The embedding diagram of the Morris-Thorne wormhole is obtained by taking a $t = \text{const.}$, $\theta = \pi/2$ section of the spherically spacetime described by the metric
\begin{equation}
    ds^{2} = - c^{2} dt^{2} + \frac{dr^{2}}{1 - b_{0}^{2}/r^{2}} + r^{2}(d\theta^{2} + \sin^{2} \theta d \phi^{2}).
    \label{MTmetric}
\end{equation}
The restricted metric, $ds^{2} = \frac{dr^{2}}{1 - b_{0}^{2}/r^{2}}+r^{2} d\phi^{2}$, can be embedded in a 3D Euclidean space with metric $ds^{2} = dz^{2} + dr^{2} + r^{2} d\phi^{2}$ such that $z = z(r)$ is the equation of the embedded surface of revolution. Thus, the metric on the surface is
\begin{equation}
    ds^{2}= \left[1 + \left(\frac{dz}{dr} \right)^{2} \right]dr^{2} + r^{2}d\phi^{2}.
    \label{embedd}
\end{equation}
It follows that
\begin{equation}
    \frac{dz}{dr}=\pm \left( \frac{1}{r^{2}/b_{0}^{2} - 1} \right)^{1/2},
    \label{dzeq}
\end{equation}
whose straightforward integration yields
\begin{equation}
    z(r) = b_{0} \cosh^{-1}\left(r/b_{0} \right),
\end{equation}
which is the equation of a catenary. Therefore the embedded section of the wormhole is a catenoid obtained by rotation of the catenary around the $z$ axis. Using the arc length of the catenary $\tau = b_{0}\sinh(z/ b_{0})$, as measured from the throat, where $r = b_{0}$ and $\tau = 0$, we get (see Eq. \eqref{cors})
\begin{equation}
    z(\tau)= b_{0}\sinh^{-1}(\tau/ b_{0}).
\end{equation}
In terms of the catenary arc length, the line element on the catenoid writes
\begin{equation}
    ds^{2} = d\tau^{2} + (\tau^{2} + b_{0}^{2}) d\phi^{2}.
    \label{catmetric}
\end{equation}

In the previous sections, we used a catenoid as a physical support for ordered nematic films whose optical metric is given either by Eq. \eqref{metric1.1} or \eqref{metric2}, depending on the liquid crystal configuration. We consider first the metric \eqref{metric1.1}, that is,
\begin{equation}
    ds^{2} = d\tau^{2} + \alpha^{2} (\tau^{2} + b_{0}^{2}) d\phi^{2}, \label{metricW}
\end{equation}
where $0 < \alpha < 1$. Now, if we assume that this is the metric of a $t = \text{const.}$, $\theta = \pi/2$ section of a wormhole in four-dimensional spacetime, we might ask what is the shape of its embedding diagram. To answer this we need to compare Eqs. \eqref{embedd} and \eqref{metricW}. Before doing it, let us write \eqref{metricW} in terms of the coordinate $r = \sqrt{\tau^{2} + b_{0}^{2}}$:
\begin{equation}
    ds^{2} = \frac{dr^{2}}{1 - b_{0}^{2}/r^{2}} + r^{2} \alpha^{2} d\phi^{2}.
    \label{met}
\end{equation}

By adjusting the length scale, making $r\alpha \to \mathcal{R}$, and $b_{0} \alpha \to \mathcal{B}_{0}$, we get
\begin{equation}
    ds^{2} = \frac{d\mathcal{R}^{2}}{\alpha^{2} \left(1 -\mathcal{B}_{0}^{2}/\mathcal{R}^{2} \right)} + \mathcal{R}^{2} d\phi^{2}.
    \label{Rmetric}
\end{equation}
Now, comparing Eqs. \eqref{Rmetric} and \eqref{embedd}, we have
\begin{equation}
    \frac{ d\mathcal{Z}}{d\mathcal{R}} = \pm \left( \frac{1}{1 - \mathcal{B}_{0}^{2}/\mathcal{R}^{2}} - \alpha^{2} \right)^{1/2},
    \label{dZeq}
\end{equation}
for the embedding in the Euclidean 3D space using cylindrical coordinates $(\mathcal{R}, \mathcal{Z}, \phi)$, where $\mathcal{Z} = \alpha z$. Equation \eqref{dZeq} reduces to \eqref{dzeq} for $\alpha = 1$. When $\mathcal{R} \to \infty$, equation \eqref{dZeq} describes a cone of opening angle $\cot^{-1} (\sqrt{1 - \alpha^{2}})$. Noting that $\mathcal{R} \geq \mathcal{B}_{0}$, the integration of Eq. \eqref{dZeq} is obtained with the coordinate transformation $\mathcal{B}_{0}/\mathcal{R} = \cos{x}$. Thus,
\begin{equation}
   \mathcal{Z}(x)= \pm \mathcal{B}_{0} \int \frac{\sqrt{1 - \alpha^{2} \sin^{2}{x}}}{\cos^{2} {x}} dx,
\end{equation}
which give us
\begin{equation}
    \begin{split}
        \mathcal{Z}(x) = \pm \mathcal{B}_{0} & \Big[ F(x,\alpha^{2})- E(x,\alpha^{2}) \\
        & + \sqrt{1 - \alpha^{2} \sin^{2}{x}}\tan x \Big],
    \end{split}
\end{equation}
where $F(x,\alpha^{2})$ and $E(x,\alpha^{2})$ are elliptic integrals of first and second kind, respectively. 

\begin{figure}[htp]
\centering
    \includegraphics[width=0.53\columnwidth]{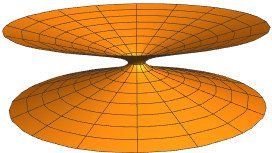}
      \includegraphics[width=0.42\columnwidth]{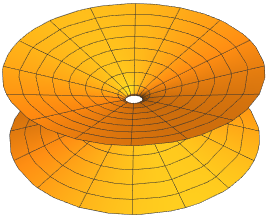}
     \caption{{The embedded surface for $0<\alpha<1$ as viewed from two different perspectives.}}
    \label{embedding}
\end{figure}

While the catenoid asymptotically tends to a plane, the surface of revolution given by $\mathcal{Z}(\mathcal{R})$ becomes a cone, as seen in Fig. \ref{embedding}. This brings us back to our Fig. \ref{cones} and to Fig. 3 of Ref. \cite{jusufi2018conical}. We see that our liquid crystal model simulates conical wormholes, like the ones associated to cosmic strings \cite{aros1997wormhole} or global monopoles \cite{jusufi2018conical}.

Equation \eqref{met} reduces to
\begin{equation}
    ds^{2} = dr^{2} + r^{2} \alpha^{2} d\phi^{2},
    \label{conemet}
\end{equation}
 when $r \to \infty$. When $0 < \alpha < 1$ (our case) this is the metric of an ordinary cone in cylindrical coordinates, in agreement with the embedding diagram shown in Fig. \ref{embedding}. For $\alpha > 1$ metric \eqref{conemet} describes a \textit{surplus} cone, a saddle-shaped surface of negative curvature. This is the case of the radial disclination configuration which is described by metric \eqref{metricW} with $\alpha \to \alpha^{-1}$. In this case, the embedding as a surface of revolution does not work since, for $\mathcal{R} > \mathcal{B}_{0}/ {\sqrt{\alpha^{2} - 1}}$, we would end up with imaginary $\mathcal{Z}$ in Eq. \eqref{dZeq} after the substitution $\alpha \to \alpha^{-1}$ . 
 
The cone metric \eqref{conemet} is a $t = \text{const.}$, $z = \text{const.}$, section of the cosmic string spacetime metric
\begin{equation}
    ds^{2} = -c^{2} dt^{2} + dz^{2} + dr^{2} + r^{2}\alpha^{2} d\phi^{2},
    \label{stringmet}
\end{equation}
as well as a $t = \text{const.}$, $\theta = {\pi/2}$, section of the global monopole spacetime metric
\begin{equation}
    ds^{2} = -c^{2} dt^{2} + dr^{2} + r^{2} d\theta^{2} + \alpha^{2} \sin\theta d\phi^{2}. 
    \label{monomet}
\end{equation}
In both cases $\alpha < 1$, meaning a deficit angle, either dihedral or solid depending on the case, is related to a positive mass distribution. Conversely, $\alpha > 1$, meaning a surplus angle, is related to a negative mass distribution, usually associated to exotic matter.

Comparing Eqs. \eqref{dzeq} and \eqref{dZeq} we see that, for $\alpha = 1$, our model describes the Morris-Thorne wormhole. For $0 < \alpha < 1$ the model with the circular disclination describes an asymptotically conical wormhole, which could be associated either to a cosmic string or to a global monopole. In contrast, the model with the radial disclination can be associated to an exotic matter string or monopole.

\section{Conclusions and perspectives}
\label{conclusions}
In this paper, we studied the propagation of light on the surface of a catenoid decorated with two different nematic liquid crystal configurations, corresponding to $+ 1$ disclinations without the core singularity. The effective optical metric found is comparable to the metrics of conical wormhole sections embedded in three-dimensional Euclidean space. In the limit where the liquid crystal film becomes isotropic, that is when the respective ordinary and extraordinary refractive indices become equal, we reproduced the trajectories of light in the Morris-Thorne wormhole 3D embedding. Otherwise, the effective geometry obtained is that of an asymptotically conical  wormhole, whose conicity could be due either to a cosmic string or to a global monopole. The trajectories, found in term of elliptic functions of first kind, are shown for some choices of  the initial conditions. The propagating wave modes were found in terms of Mathieu functions and are depicted both as functions of the position or as intensity profiles on the catenoid. 

The radial disclination on the catenoid deserves further investigation since it may become a valuable analogue model for wormholes involving exotic cosmic strings. As suggested in Refs.  \cite{visser1989traversable2} and \cite{cramer1995natural}, negative mass cosmic strings wrapped around primordial wormholes would stabilize and therefore permit them to survive up to present time. The strings would act as struts impeding the wormhole mouth to close. Furthermore, an issue for future research to explore is the relation between the propagation of light studied here and bouncing tachyons \cite{fredenhagen2003minisuperspace} which obey essentially Eq. \eqref{Mathieu}. Also, Fig. \ref{fig:effpot} (a) suggests the possibility of bound states, something that should be further investigated. In particular, cases with $m \neq 0$. Finally, a refinement of our model can be done by the inclusion of a geometry-induced potential \cite{schultheiss2020light} in the wave equation, as done in Ref. \cite{dandoloff2010geometry} for the Schr\"odinger equation on the catenoid.

\acknowledgements
F.S.A. is grateful to CNPq for a PDJ scholarship, J.D.M.L. thanks FACEPE for an IBPG scholarship, A.P.S. and F.M. thank FACEPE, CNPq, and CAPES for financial support.

\bibliographystyle{apsrev4-2}
\bibliography{main}
\end{document}